\documentclass[aps,pra,twocolumn,groupedaddress,
showpacs]{revtex4}

\usepackage{graphics}
\usepackage{epsfig}
\usepackage{amsmath}

\begin{document}

% number in the upper righthand corner of the title page in preprint
% mode.
%\preprint{}

\title{Quantum limited force measurement in a cavityless optomechanical 
system}


\author{Rachele Fermani}
%\email{}
%\thanks{}
\affiliation{INFM \& Dipartimento di Fisica, Universit\`{a} di Camerino,
I-62032 Camerino, Italy.}
\author{Stefano Mancini}
%\email{}
%\thanks{}
\affiliation{INFM \& Dipartimento di Fisica, Universit\`{a} di Camerino,
I-62032 Camerino, Italy.}
\author{Paolo Tombesi}
%\email{}
%\thanks{}
\affiliation{INFM \& Dipartimento di Fisica, Universit\`{a} di Camerino,
I-62032 Camerino, Italy.}

\date{\today}

\begin{abstract}
We study the possibility of revealing 
a weak coherent force by using a pendular 
mirror as a probe, and coupling this to a radiation field, 
which acts as the meter, 
in a cavityless configuration.
We determine the sensitivity of such a scheme 
and show that the use of an entangled meter state
greatly improves the ultimate detection limit.
We also compare this scheme with that 
involving an optical cavity.
\end{abstract}


\maketitle

Optomechanical systems play a crucial role in a variety 
of precision measurement
like gravitational wave detection \cite{GRAV} and atomic force 
microscope \cite{AFM}.
They are based on the interaction between a movable mirror, 
a {\it probe} experiencing tiny forces, 
and a radiation field, a {\it meter} reading out the mirror's position.  

In these applications one needs very high resolution  
measurements and a good control of the various noise sources, 
because one has
to detect the effect of very weak forces. 
In the simpler setup \cite{SETUP}, 
optomechanical systems are usually intended 
as a Fabry-Perot cavity with a movable end 
mirror coupled to the external force and to the radiation probe.
As shown by the pioneering work
of Braginsky \cite{BRAG}, even though all classical noise sources
had been minimized, the detection of weak forces in such optomechanical
systems would ultimately be
determined by quantum fluctuations and the Heisenberg 
uncertainty principle.
In particular, quantum noise has two fundamental sources, 
the photon shot noise 
of the laser beam,  
and the fluctuations
of the mirror position due to radiation pressure.
The two quantum noises determine the so-called
{\em standard quantum limit} (SQL). 
It has been argued that the use of squeezed meter
state allows to overcome the SQL \cite{CAV80}.

Here, differently from the standard optomechanical setup,
\cite{SETUP}, we do not consider an optical cavity but 
only a single mirror, illuminated by an intense and highly 
monochromatic optical beam. 
We shall derive the SQL for such a system and we shall 
show that it could be beaten by using entangled meters. 

\begin{figure}
\begin{center}
\includegraphics[width=0.3\textwidth]{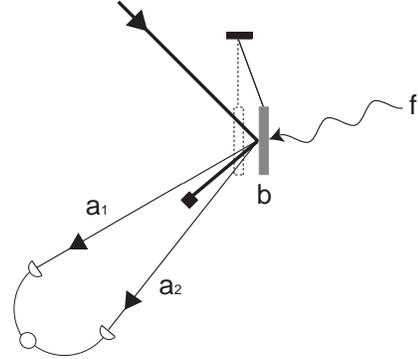}
\end{center}
\caption{\label{fig1} 
Schematic description of the studied system. A laser field at frequency
$\omega_{0}$ impinges on the mirror vibrating at frequency $\Omega$.
In the reflected field two sideband modes are excited at
frequencies $\omega_{1}=\omega_{0}-\Omega$ and
$\omega_{2}=\omega_{0}+\Omega$.}
\end{figure}

We consider a perfectly reflecting mirror and an intense 
quasi-monochromatic
laser beam impinging on its surface (see Fig.~\ref{fig1}).
The laser beam
is linearly polarized along the mirror surface and focused in such a way 
as to
excite Gaussian acoustic modes of the mirror. These modes describe small
elastic deformations of the mirror along the direction orthogonal to its
surface and are characterized by a small waist, a large quality factor 
and a
small effective mass \cite{PIN99}. It is possible to adopt a single
vibrational mode description limiting the detection bandwidth to include a 
single
mechanical resonance of frequency $\Omega$.
In this description the incident
laser beam, with frequency $\omega_{0}$, is reflected into an elastic 
carrier
mode, with the same frequency $\omega_{0}$, and two additional weak 
anelastic
sideband modes with frequencies $\omega_{0}\pm\Omega$ \cite{PRA1203}. 
The physical process is
very similar to a stimulated Brillouin scattering, even though in this 
case
the Stokes and anti-Stokes component are back-scattered by the acoustic 
wave
at reflection, and the optomechanical coupling is provided by the 
radiation
pressure. Treating classically the intense incident beam (and the carrier
mode), the quantum system is composed of three interacting quantized 
bosonic
modes, i.e. the vibrational mode and the two sideband modes. In our
description, vibrational, Stokes and anti-Stokes modes are 
characterized by ladder operators $\hat b$, $\hat a_{1}$ and $\hat a_{2}$ respectively.
In \cite{PRA1203} an effective
interaction Hamiltonian for that system has been derived as
\begin{equation}\label{Heff}
{\hat{H}}_{eff}=
-i\hbar\chi\left({\hat{a}}_{1}{\hat{b}}-{\hat{a}}_{1}^{\dag
}{\hat{b}}^{\dag}\right)
-i\hbar\theta\left({\hat{a}}_{2}{\hat{b}}^{\dag}-{\hat
{a}}_{2}^{\dag}{\hat{b}}\right)\,, 
\end{equation}
where $\chi$ and $\theta$ are couplings constants 
proportional to $\sqrt{\wp}$, with $\wp$ the incident laser power.
Their ratio
$\theta/\chi=\left[
(\omega_{0}+\Omega)/(\omega_{0}-\Omega)\right]  ^{1/2} \geq1$ only
depends on the involved frequencies. The system dynamics is
satisfactorily reproduced by the Hamiltonian of
Eq.~(\ref{Heff}) as long as the dissipative coupling of the
mirror vibrational mode with its environment is negligible. This
happens if the interaction time is much smaller than the
relaxation time of the vibrational mode 
and therefore it means having a high quality factor 
vibrational mode \cite{TIT99}.

We now consider the action of a classical coherent force on the
probe and its 
readout through radiation fields.
Since Eq.(\ref{Heff}) is already written
in a frame rotating at frequency $\Omega$ 
\cite{PRA1203}, if the force is constant,
the total Hamiltonian would be
\begin {equation}\label{H}
     \hat H=\hat H_{eff}-\hbar \Omega f\left(\hat b e^{-i\Omega t}
     +\hat b^{\dagger} e^{i\Omega t}
     \right),
\end{equation}
where $f$ denotes the dimensionless force strength. 
Eq.(\ref{H}) leads to the following set of Heisenberg equations
\begin{subequations}\label{eqsa1ba2}
\begin{eqnarray}
    \dot{\hat a}_{1}&=&\chi {\hat b}^{\dag}\,,
    \label{eqa1} \\
    \dot{\hat b}&=&\chi {\hat a}_{1}^{\dag}-\theta {\hat 
    a}_{2}+i\Omega f e^{i \Omega t}\,,
    \label{eqb} \\
    \dot {\hat a}_{2}&=&\theta {\hat b}
    \label{eqa2} \,,
\end{eqnarray}
\end{subequations}
whose solutions read
\begin{subequations}\label{a1tbta2t}
\begin{eqnarray}
    &&{\hat a}_{1}(t)=\frac{1}{\Theta^{2}}\left[\theta^{2}-\chi^{2}
    \cos\left(\Theta t\right)\right]{\hat a}_{1}(0)
    +\frac{\chi}{\Theta}\sin\left(\Theta t\right) {\hat b}^{\dag}(0)
    \nonumber\\
    &&-\frac{1}{\Theta^{2}}\left[\chi\theta-\chi\theta
    \cos\left(\Theta t\right)\right]{\hat a}_{2}^{\dag}(0)
    \nonumber\\
    &&-\left[ \Omega \sinh \left( \Theta t\right)
    -i \left( e^{-i \Omega t}-\cos \left( \Theta t\right) \right) 
    \right] 
    \frac{\Omega}{\Omega^{2} -\Theta^{2}} \chi f\,,
    \label{a1t}
    \\
    &&{\hat b}(t)=\frac{\chi}{\Theta}\sin\left(\Theta t\right)
    {\hat a}_{1}^{\dag}(0)
    +\cos\left(\Theta t\right) {\hat b}(0)
    \nonumber\\
    &&-\frac{\theta}{\Theta}\sin\left(\Theta t\right) {\hat a}_{2}(0)
    \nonumber \\
    &&- \left[ \Omega \cos \left( \Theta t\right) +i \Theta \sin \left(
    \Theta t\right) + \Omega e^{i \Omega t} \right] 
    \frac{\Omega}{\Omega^{2} - \Theta^{2}} f 
    \,,
    \label{bt}
    \\
    &&{\hat a}_{2}(t)=\frac{1}{\Theta^{2}}\left[\chi\theta-\chi\theta
    \cos\left(\Theta t\right)\right]{\hat a}_{1}^{\dag}(0)
    +\frac{\theta}{\Theta}\sin\left(\Theta t\right) {\hat b}(0)
    \nonumber\\
    &&-\frac{1}{\Theta^{2}}\left[\chi^{2}-\theta^{2}
    \cos\left(\Theta t\right)\right]{\hat a}_{2}(0)
    \nonumber\\
    &&-\left[ \Omega \sin \left( \Theta t\right) + i \left( 
    e^{i \Omega t} - \cos \left( \Theta t\right) \right) \right] 
    \frac{\Omega}{\Omega^{2}-\Theta^{2}} \theta f \,,
    \label{a2t}
\end{eqnarray}
\end{subequations}
with $\Theta=\sqrt{\theta^{2}-\chi^{2}}$.

We consider the mirror initially in a thermal state \cite{QOPT}
\begin{equation}\label{rhob}
    \hat \rho_{b}=\frac{1}{1+{\cal N}_{th}} \sum_{n}
    \left(\frac{{\cal N}_{th}}{1+{\cal 
    N}_{th}}\right)^{n} |n\rangle \langle n|\,,
\end{equation}
where the mean number of thermal excitations is given by
\begin{equation}\label{Nth}
             {\cal N}_{th}=\frac{1}{e^{\hbar \Omega/
	     k_{B} T}-1}\,.
\end{equation}
with $k_{B}$ the 
Boltzmann constant and $T$ the equilibrium temperature.
Furthermore, quite generally, we assume the meter modes initially in a 
pure state of the form
\begin{equation}\label{Psi12}
|\Psi\rangle_{12}=\sqrt{1-\tanh^{2}s}\sum_{n=0}^{\infty}
\left(\tanh s\right)^n |n\rangle_{1}\,
|n\rangle_{2}\,,
\end{equation}
with the parameter $s\in{\bf R}$.
This state for $s=0$ gives the tensor product of vacuum state 
for the two sideband modes, instead, if $s\ne 0$ it represents
a two mode squeezed state \cite{QOPT}, which shows entanglement. 
In that case the two meter modes ${\hat{a}}_{1}$ and ${\hat{a}}_{2}$ 
are not initially empty, thus, they could in turn excite their own sideband 
modes with coupling costants say $\chi_{1,2}$ and $\theta_{1,2}$. 
The latter are proportional to the corresponding powers $\wp_{1,2}$ 
of the modes ${\hat{a}}_{1}$, ${\hat{a}}_{2}$ determined by 
$\sinh^{2} s$ \cite{QOPT}. 
We assume that $\wp \gg \wp_{1,2}$, hence 
$\chi,\, \theta  \gg  \chi_{1,2},\,\theta_{1,2}$,   
so that in Eq.(\ref{Heff}) we neglected terms 
related to $\chi_{1,2}$ and $\theta_{1,2}$. 
However, to not neglect the force term in Eq.(\ref{H}), we also require 
the condition $\Omega f \gg \chi_{1,2},\,\theta_{1,2}$ to be satisfied.

Let us now consider the heterodyne detection \cite{YS} on the reflected 
sideband modes ${\hat{a}}_{1}$, ${\hat{a}}_{2}$. That is, we consider the possibility to 
simultaneously measure the real and the imaginary part of the operator
\begin{equation}\label{Zvarphi}
    \hat Z_{\varphi}=\hat a_{1}e^{i\varphi}
    -\hat a_{2}^{\dagger}
    e^{-i\varphi}\,,
\end{equation}
where $\varphi$ is a phase that can be experimentally adjusted.
We now introduce the amplitude and phase quadratures
for the two optical modes as $\hat X_{j}=(\hat a_{j}+\hat 
a_{j}^{\dag})/{2}$, $\hat Y_{j}=-i(\hat a_{j}-\hat 
a_{j}^{\dag})/{2}$, $j=1,2$, with commutation relation 
$[\hat X_{j},\hat Y_{k}]=\delta_{jk}/2$.
Then, it is well known \cite{QOPT}
that the state (\ref{Psi12}) with $s\ne 0$ allows 
reduced variances for
$\hat X_{1}-\hat X_{2}$ and $\hat Y_{1}+\hat Y_{2}$
with respect to the case of $s=0$.
This could be fruitfully 
exploited to reduce the noise on the readout. 
Inspired by this fact 
we choose $\varphi=\pi$ in Eq.(\ref{Zvarphi}) and obtain
\begin{equation}\label{Zpi}
    \hat Z_{\pi}=-\left(
    \hat X_{1}-\hat{X}_{2}\right)
    -{i}\left(\hat Y_{1}+
    \hat Y_{2}\right)\,.
\end{equation}
More specifically, we shall consider only the imaginary part of $\hat 
Z_{\pi}$, which, by virtue of Eqs.(\ref{Zpi}) and (\ref{a1tbta2t}), 
results
\begin{eqnarray}\label{ZI}
    &&\hat Z_{I}(t) =
    \frac{\left(\theta-\chi\right)}{\Theta^{2}}
    \left[\theta+\chi\cos\left(\Theta t\right)\right]\hat Y_{1}(0)
    \nonumber\\
    &&+\frac{\left(\theta-\chi\right)}{\Theta^{2}}
    \left[\chi+\theta\cos\left(\Theta t\right)\right]\hat Y_{2}(0)
    \nonumber\\
    &&+\frac{\left(\theta-\chi\right)}{\Theta}\sin\left(\Theta 
    t\right) \left[\frac{\hat b(0)-\hat b^{\dag}(0)}{2i}\right]
    \nonumber\\
    &&+\left(\theta-\chi\right) \left[ \cos \left( \Theta t\right)-
    \cos \left( \Omega t\right) \right] \frac{\Omega}{\Omega^{2}-
    \Theta^{2}} f
    \,.
\end{eqnarray}
Then, the signal, using Eqs.(\ref{rhob}) and (\ref{Psi12}), will be
\begin{equation}\label{calS}
    \langle Z_{I}\rangle  \equiv {\cal S} f
    =\left(\theta-\chi\right) \left[ \cos \left( \Theta t\right)-
    \cos \left( \Omega t\right) \right] \frac{\Omega}{\Omega^{2}-
    \Theta^{2}} f
   \,,
\end{equation}
while the noise
\begin{eqnarray}\label{calN}
  &&\langle Z_{I}^{2}\rangle-\langle Z_{I}\rangle^{2} \equiv {\cal N}  
  \nonumber\\ 
  &&=\frac{-1}{4\Theta^{4}}
         	 \bigg[2 \theta \chi \left( \theta^{2}+\chi^{2} \right) 
	 \left( 1- \cos \left( \Theta t\right) \right)^{2} +
	 8 \theta^{2} \chi^{2} \cos \left( \Theta t\right) 
	 \nonumber \\
	 &&-
	 \left( \theta^{2}+\chi^{2} \right)^{2} \left( 1+ 
	 \cos^{2} \left( \Theta t\right) \right) \bigg] 
	 \left( 1+ 2 \sinh^{2} (s) \right) 
	 \nonumber\\
	 &&-\,\frac{1}{2 \Theta^{4}} 
	 \bigg[-\left( \theta^{2} +\chi^{2} \right)^{2} 
	 \cos \left( \Theta t\right) + 2\theta^{2} \chi^{2} 
	 \left( 1+ \cos^{2} \left( \Theta t\right) \right) 
	 \nonumber\\
	  &&-  
	 \theta \chi \left( \theta^{2} +\chi^{2}\right) 
	 \left( 1 - \cos \left( \Theta t\right) \right)^{2} \bigg]
	 \left( \sinh (2 s) \right) 
	 \nonumber \\
	 &&+
	 \frac{\left( \theta - \chi\right)^{2}}{4 \Theta^{2}}	
	 \sin^{2} \left( \Theta t\right)
	 \left( 1 +2 {\cal N}_{th} \right).  
	 \end{eqnarray}

The relevant quantity determining the sensitivity of the system
is the signal to noise ratio (SNR)
\begin{equation}\label{calR}
{\cal R}=\frac{|{\cal S}|}{\sqrt{{\cal N}}}f\,,
\end{equation}
which must be $\ge 1$ to reveal the force. Hence, the condition 
${\cal R}=1$ gives the minimum detectable force, i.e.
\begin{equation}
f_{min}=\frac{\sqrt{{\cal N}}}{|{\cal S}|}\,.
\end{equation}

We can see that the thermal noise increases the values assumed 
by $f_{min}$ for all times, but for $\Theta t=\pi$ it gives no contribution. 
At that time we have
\begin{equation}\label{fmin}
    f_{min}(\Theta t=\pi)=\frac{\Omega^{2}-\Theta^{2}}{\theta+\chi}
    \frac{e^{-s}}{\sqrt{2} \Omega \left( 1-\cos 
    \left( \pi \Omega/\Theta\right) \right)}\,,
    \end{equation}
which shows the possibility to improve the sensitivity 
by using entanglement (provided $\Omega/(2\Theta)$ not integer).

Now, we compare our model with that involving a single 
mode optical cavity. We consider an intense laser beam (of power $\wp$)
exciting a single cavity mode and realizing, in the semiclassical 
approximation the effective Hamiltonian \cite{EPL03} 
\begin{equation}\label{Heffprime}
\hat H_{eff}=-\hbar\Delta\hat a^{\dag}\hat a
+\hbar \Omega \hat b^{\dag} \hat b
+\hbar g\left(\alpha^{*}\hat a 
+\alpha \hat a^{\dag}\right)
\left(\hat b  +  \hat b^{\dag}\right)\,,
\end{equation}
where $\alpha\propto\sqrt{\wp}$ denotes the classical cavity field 
amplitude,
$\Delta$ the cavity detuning, $g$ the optomechanical coupling,  
and $\hat{a}$ the radiation field quantum fluctuations.
The meaning of the other symbols is the same as in the main text.
By adding the driving $-\hbar\Omega f\left(\hat b + \hat 
b^{\dag}\right)$ to Eq.(\ref{Heffprime}), we are led to the following 
Heisenberg equations
\begin{subequations}\label{lineqs}
\begin{eqnarray}
    \dot{\hat{a}}&=&i\Delta\hat{a}- ig\alpha
    \left(b+b^{\dag}\right)\,,
    \\
    \dot{\hat{b}}&=&-i\Omega \hat{b}
    -ig \left(\alpha^{*}\hat{a}+\alpha\hat{a}^{\dag}\right)
    +i\Omega f\,,
\end{eqnarray}
\end{subequations}
Choosing $\Delta=0$, $\alpha \in {\bf R}$, and introducing the 
field quadratures 
${\hat X}=(\hat{a}+\hat{a}^{\dag})/{2}$,
${\hat Y}=-i(\hat{a}-\hat{a}^{\dag})/{2}$,
we immediately see that only the phase quadrature $\hat{Y}$
carries out information about the force.
As a matter of fact, from Eqs.(\ref{lineqs}), we obtain
\begin{eqnarray}\label{Yt}
   && \hat{Y}(t) = -
    \frac{g\alpha}{\Omega}\sin\left(\Omega t\right)
    \left(\hat{b}(0)+\hat{b}^{\dag}(0)\right)
    \nonumber\\
    &&+i\frac{g\alpha}{\Omega}
    \left[1-\cos\left(\Omega t\right)\right]
    \left(\hat{b}(0)-\hat{b}^{\dag}(0)\right)
    \nonumber\\
    &&+\left(\frac{2 g\alpha}{\Omega}\right)^{2}\left[\Omega t
    -\sin\left(\Omega t\right)\right]
    \hat{X}(0)
    \nonumber\\
    &&+\frac{2g\alpha}{\Omega}\left[\Omega t-
    \sin\left(\Omega t\right)\right]f
    +\hat{Y}(0)\,.
\end{eqnarray}
As the initial state we consider the thermal state (\ref{rhob})
for the probe and a squeezed state for the meter, i.e.
$\exp[\zeta^{*}(\hat{a})^{2}-\zeta({\hat a}^{\dag})^{2}]|0\rangle$
with $\zeta=(s/2)\exp(2i\phi)$ the squeezing parameter \cite{QOPT}.

Since we are considering the good cavity limit \cite{Note}, 
we restrict our analysis to only the cavity mode and suppose to be 
able to perform its homodyne detection \cite{HOM}. 
One then gets the following signal
\begin{equation}\label{sig}
    \langle\hat{Y}(t)\rangle
    \equiv {\cal S}(t)\, f
    =\frac{2g\alpha}{\Omega}\left[
    \Omega t-\sin\left(\Omega t\right)\right]\,f\,.
\end{equation}
The corresponding noise can be calculated by means of 
the initial thermal state of the probe, 
and the squeezed state for the meter.
Thus, we get
\begin{eqnarray}\label{noi}
    &&\left\langle\hat{Y}^{2}(t)\right\rangle-\langle \hat Y(t)\rangle^{2}
    \equiv{\cal N}(t)
    \\
    &&=\left(\frac{g\alpha}{\Omega}\right)^{2}
    \sin^{2}\left(\Omega t\right) 
    \left({1}+2{\cal N}_{th}\right)
    \nonumber\\
    &&+\left(\frac{g\alpha}{\Omega}\right)^{2}
    \left[1-\cos\left(\Omega t\right)\right]^{2}
    \left({1}+2{\cal N}_{th}\right)
    \nonumber\\
    &&+\left(2\frac{g\alpha}{\Omega}\right)^{4}
    \left[\Omega t-\sin\left(\Omega t\right)\right]^{2}
    \left[e^{-2s}\cos^{2}\phi+e^{2s}\sin^{2}\phi\right]
    \nonumber\\
    &&+\left[e^{-2s}\sin^{2}\phi+e^{2s}\cos^{2}\phi\right]
    \nonumber\\
     &&+2\left(\frac{2g\alpha}{\Omega}\right)^{2}
    \left[\Omega t-\sin\left(\Omega t\right)\right]
    \left[e^{2s}-e^{-2s}\right]\sin\phi\cos\phi
    \,.
    \nonumber
\end{eqnarray}
It is easy to see that at times such that 
$\Omega t=2\pi$ the thermal noise does not contribute.
Then, minimizing over $\phi$ we get for such times
\begin{equation}\label{Nmin}
    {\cal N}_{min}(\Omega t=2\pi)=
    \left[1+4\pi^{2}\left(\frac{2g\alpha}{\Omega}\right)^{4}
    \right]e^{-2s}\,,
\end{equation}
that is 
\begin{equation}\label{fmin2}
f_{min}(\Omega t=2\pi)= \frac{\left[
1+4\pi^{2}\left(2g\alpha/\Omega\right)^{4}\right]^{1/2}}
{4\pi\left(g\alpha/\Omega\right)}e^{-s}\,.
\end{equation}

\begin{figure}
\begin{center}
\includegraphics[width=0.35\textwidth]{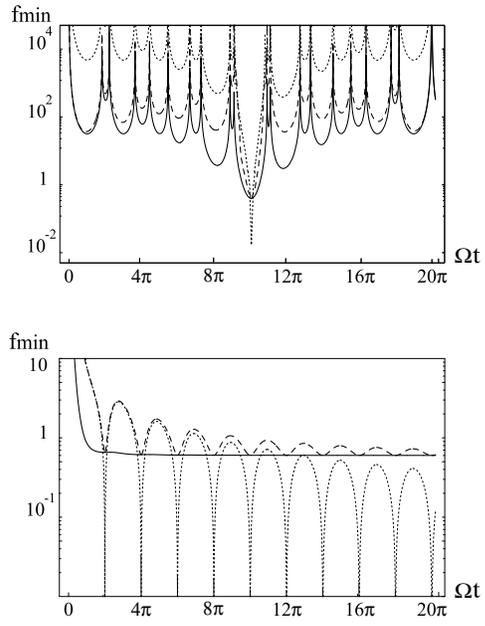}
\end{center}
\caption{\label{fig2} 
Log plot of the minimum detectable force versus scaled time $\Omega t$, 
for the cavityless model (top plot), and for the cavity model (bottom plot).
In each plot the continous line represents 
the SQL; the dashed line refers to ${\cal N}_{th}=300$ and $s=0$; 
the dotted line refers to ${\cal N}_{th}=300$ and $s=5$.
The values of other parameters are: $\theta/\chi=1.025$, 
$\Omega/\Theta \simeq 10$, $g \alpha / \Omega = 0.2$.}
\end{figure}

The minimum detectable force, for each of the models, 
is shown in Fig.\ref{fig2}
as a function of time for different values of $s$ and ${\cal N}_{th}$. 
In both cases $f_{min}$ for ${\cal N}_{th}=s=0$
represents the SQL. 
For the cavityless model (top plot), we notice oscillations 
due to the presence of two different 
frequencies $\Theta$, $\Omega$ in Eq.(\ref{calS}), while for 
the cavity model (bottom plot) 
the oscillations only depend on the mirror frequency $\Omega$. 
We can see that for all times in which the mirror is disentangled 
from the radiation, Eq. (\ref{fmin}) 
and Eq.(\ref{fmin2}) show the possibility to go beyond the SQL. 
Note that Fig. \ref{fig2} represents only a ``qualitative'' comparison 
between the two models.

Looking at the scaling in terms of the laser power, in both 
Eqs. (\ref{fmin}) and (\ref{fmin2}) we recognize 
the same contributions: $\frac{1}{\sqrt{\wp}}$ due to the 
shot noise (prevealing at small laser power), and $\sqrt{\wp}$ 
due to the radiation pressure noise (prevealing at large laser power). 
Anyway the two contributions are combined in a different manner; 
for the optical cavity model we have 
$f_{min} \propto  \sqrt{\frac{1}{\wp}+\wp}$, while in the cavityless 
optomechanical model we have 
$f_{min} \propto \frac{1}{\sqrt{\wp}}-\sqrt{\wp}$. 
In the latter case, it seems that $f_{min}$ could be reduced to $0$ by simply 
choosing a proper value of $\wp$. 
This could happen for $\Omega^{2}-\Theta^{2}=0$, which however 
invalidates our model, 
since the Hamiltonian (\ref{Heff}) has been derived in the limit 
$\Omega^{2} \gg \Theta^{2}$ \cite{PRA1203}.  
Moreover $f_{min}$ cannot go to $0$, because of the 
condition $\Omega f \gg \chi_{1,2},\, \theta_{1,2}$ 
which requires a large enough frequency $\Omega$ in order to beat the SQL 
(this condition holds e.g. for the parameters values of Fig.\ref{fig2}).  

It should be noted that in the cavityless model the time is essentially 
set by the incident optical pulse length. Therefore, it is 
particularly suited to perform pulsed measurement on the probe, 
while the cavity 
model presupposes a stationary condition between meter and probe, hence 
measurement on a long time scale. 
In both cases the use of a nonclassical meter state allows improvement 
of the performances only for particular interaction times, while the use of 
a nonclassical probe state would allow such improvement for
almost all times \cite{HOL79,EPL03}.

In conclusion, we have presented a cavityless optomechanical model 
to reveal weak coherent forces and we have compared it with a cavity one. 
In particular, we have shown that nonclassical meter 
states allow one to beat the SQL
greatly improving the sensitivity. 
Thus, we have shown that entanglement in a cavityless optomechanical scheme 
plays almost the same role as does squeezing in one which use a cavity. 
Finally, the cavityless optomechanical model could be useful in a number 
of applications,
especially those involving micro-opto-mechanical-sensors \cite{MOMS}.

\section*{Acknowledgments}
We gratefully acknowledge several discussions with David Vitali.


\begin{references}

\bibitem{GRAV}
A. Abramovici, {\it et al.},  Science {\bf 256}, 325 (1992);
R. Loudon, Phys. Rev. Lett. {\bf 47}, 815 (1981).

\bibitem{AFM}
J. Mertz, {\it et al.}, Appl. Phys. Lett. {\bf 62}, 2344 (1993);
T. D. Stowe, {\it et al.}, Appl. Phys. Lett. {\bf 71}, 288 (1997).

\bibitem{SETUP}
S. Mancini and P. Tombesi,
Phys. Rev. A {\bf 49}, 4055 (1994);
C. Fabre, {\it et al.}, Phys. Rev. A 
{\bf 49}, 1337 (1994);
G.J. Milburn, K. Jacobs, D. F. Walls 
Phys. Rev. A {\bf 50}, 5256 (1994);
C. K. Law, Phys. Rev. A {\bf 51}, 1537 (1995);
S. P. Vyatchanin and A. B. Matsko, 
J. Exp. Theor. Phys. {\bf 82}, 1007 (1996);
{\bf 83}, 690 (1996);
S. Mancini, V. I. Man'ko and P. Tombesi,
Phys. Rev. A {\bf 55}, 3042  (1997);
S. Bose, K. Jacobs and P. L. Knight,
Phys. Rev. A {\bf 56}, 4175 (1997);
K. Jacobs, {\it et al.},
Phys. Rev. A {\bf 60}, 538 (1999);
V. Giovannetti and D. Vitali,
Phys. Rev. A {\bf 63}, 023812 (2001). 
D. Vitali, {\it et al.},
Phys. Rev. A {\bf 65}, 063803 (2002).

\bibitem{BRAG}
V. B. Braginsky and F. Ya Khalili, 
{\it Quantum Measurements}, 
(Cambridge University Press, Cambridge, 1992).

\bibitem{CAV80}
C. M. Caves, Phys. Rev. Lett. {\bf 45}, 75 (1980);
R. S. Bondurant and J. H. Shapiro, Phys. Rev. D {\bf 30}, 
2548 (1984);
A. F. Pace, M. J. Collett and D. F. Walls,
Phys. Rev. A {\bf 47}, 3173 (1993).

\bibitem{PIN99}
M. Pinard, {\it et al.},
Eur. Phys. J. D {\bf 7}, 107 (1999).

\bibitem{PRA1203}
S. Pirandola, S. Mancini, D. Vitali, P. Tombesi, Phis. Rev. A 
{\bf 68}, 062317 (2003). 

\bibitem{TIT99}
I. Tittonen, {\it et al.},
Phys. Rev. A {\bf 59}, 1038 (1999).

\bibitem{QOPT}
D. F. Walls and G. J. Milburn, 
{\it Quantum Optics},
(Springer, Berlin, 1994).

\bibitem{YS}
H. P. Yuen and J. H. Shapiro,
IEEE Trans. Info. Theory {\bf IT-26}, 78 (1980).

\bibitem{MOMS}
A. N. Cleland and M. L. Roukes, Nature(London)
{\bf 392}, 168 (1998);
H. J. Mamin and D. Rugar, 
Appl. Phys. Lett. 
{\bf 79}, 3358 (2001).

\bibitem{HOL79}
J. N. Hollenhorst, Phys. Rev. D {\bf 19}, 1669 (1979).

\bibitem{EPL03}
S. Mancini and P. Tombesi,
Europhys. Lett. {\bf 61}, 8 (2003).

\bibitem{Note}
Notice that the model of Eq.(\ref{Heff}) cannot be trivially 
obtained as the bad cavity limit of Eq.(\ref{Heffprime}), because the 
two Hamiltonians have been derived under different assumptions, see 
e.g. Refs. \cite{SETUP, PRA1203}. 

\bibitem{HOM}
H. M. Wiseman and G. J. Milburn, Phys. Rev. A {\bf 47},
642 (1993).





\end{references}
\end{document}